\documentclass[twocolumn,american,aps,showpacs]{revtex4}
\usepackage[latin9]{inputenc}
\usepackage{amsmath}
\usepackage{amssymb}
\usepackage{graphicx}
\usepackage{esint}

\makeatletter
\@ifundefined{textcolor}{}
{%
 \definecolor{BLACK}{gray}{0}
 \definecolor{WHITE}{gray}{1}
 \definecolor{RED}{rgb}{1,0,0}
 \definecolor{GREEN}{rgb}{0,1,0}
 \definecolor{BLUE}{rgb}{0,0,1}
 \definecolor{CYAN}{cmyk}{1,0,0,0}
 \definecolor{MAGENTA}{cmyk}{0,1,0,0}
 \definecolor{YELLOW}{cmyk}{0,0,1,0}
}

\usepackage{dcolumn}
\usepackage{bm}
\usepackage{wrapfig}
\usepackage{subfigure}
\usepackage{ucs}
\usepackage{lineno}
\usepackage{epsfig}
\usepackage{psfrag}\usepackage{epstopdf}
\DeclareGraphicsExtensions{.pdf,.eps,.png,.jpg,.mps}

\usepackage{babel}

\usepackage{babel}

\makeatother

\usepackage{babel}
\begin{document}

\title{Quantum entanglement in the neighborhood of pseudo-transition for
a spin-1/2 Ising-XYZ diamond chain}

\author{ I. M. Carvalho$^{1}$, J. Torrico$^{2}$, S. M. de Souza$^{1}$,
M. Rojas$^{1}$ and O. Rojas$^{1}$. }

\affiliation{$^{1}$Departamento de Física, Universidade Federal de Lavras, 37200-000,
Lavras-MG, Brazill}

\affiliation{$^{2}$Instituto de Ciências Exatas, Universidade Federal de Alfenas,
37133-840, Alfenas-MG, Brazil}
\begin{abstract}
Recently has been observed for some one-dimensional models that exhibit unexpected
pseudo-transitions and quasi-phases.  This pseudo-transition resembles
a first- and second-order phase transition simultaneously.  One of
those models is the spin-1/2 Ising-XYZ diamond chain, composed of
Ising spin particles at the nodal sites and the Heisenberg spin particles
at the interstitial sites.  Where we assume Ising-type interaction
between the nodal and interstitial sites, the Heisenberg-type interaction
between interstitial sites, and with an external magnetic field applied
along the $z$-axis.  This model presents an exact analytical solution
applying the transfer matrix technique, which shows 3 phases at zero
temperature in the vicinity of pseudo-transition.  The pseudo-transition
separates quasi-phases, these quasi-phases still hold at a finite
temperature most of the pattern configurations of a true phase at
zero temperature.  Here we study the quantum entanglement of pair
spin particles in the quasi-phase regions, which can be measured through
the concurrence. Then we observe an unexpected behavior in the concurrence,
that is below pseudo-critical temperature the concurrence remains
almost constant up to pseudo-critical temperature, but above the pseudo-critical
temperature, the concurrence behaves as for the standard one-dimensional
spin models.   Further, we consider the entropy behavior of the system,
below pseudo-critical temperature the entropy becomes almost null,
while above pseudo-critical temperature the system exhibits standard
behavior as for ordinary one-dimensional spin models.
\end{abstract}

\maketitle

\section{Introduction}

The entanglement is one of the most emblematic properties of quantum
systems, which have been investigated in several fields, particularly
in quantum information\cite{Ben1,Ben2,bennett1,lamico}, quantum computing\cite{loss,ben},
among others. Quantum entanglement in spin systems is a relevant emerging
field. In fact, spin chains are suitable candidates for generation
and manipulation of entanglement\cite{divi,bose}. The Heisenberg
spin chain has been suggested to construct a quantum computer in many
physical systems such as quantum dots\cite{bur}, nuclear spin\cite{kane},
electronic spin\cite{vri}, superconductor\cite{nishi}, and optical
lattice\cite{duan}.

In the context of the spin chain with diamond structure, the Ising-Heisenberg
diamond chain can describe the features of some real compounds, such
as Cu$_{3}$(CO$_{3}$$)_{2}$(OH)$_{2}$, known as the mineral azurite\cite{kiku},
which has attracted a great deal of attention. Besides, the magnetic
properties of the mineral likasite Cu$_{3}$(OH)$_{5}$(NO$_{3}$)·2H$_{2}$O\cite{kikuchi},
that for some conditions can be described as a diamond chain.

Recently, several studies have been achieved concerning the thermal
entanglement in spin diamond chains\cite{ono-1,hov,pere,Gao,mrojas}.
The quantum teleportation through a couple of Ising-XXZ diamond chain
has also been examined in detail in reference \cite{moi}. Furthermore,
the quantum correlation has been quantified by the trace distance
discord to describe the quantum critical behaviors in the Ising-XXZ
diamond structure at finite temperature\cite{cheng}.

Despite the absence of finite-temperature phase transitions in one-dimensional
models, some peculiar models exhibit the so-called finite-temperature
pseudo-transition, as discussed by Ferrari and Russo \cite{ferrari}.
Using the formalism of the microcanonical ensemble to a unidimensional
chain, where a critical region was observed exhibiting an unusual
behavior of insensitivity due to thermal excitation, leading to a
pseudo-phase transition.\cite{ferrari}. On the other hand, the evidence
of pseudo-transition and quasi-phases were also observed in canonical
ensemble formalism\cite{timo}.

More recently, were also observed this unusual behavior in one-dimensional
models with some peculiar chain structures which exhibit a quasi-phases
and pseudo-transitions\cite{qphases}. This evidence has been observed
in the following models: the Ising-XYZ diamond chain structure in
an external magnetic field\cite{rojas}, the double-tetrahedral chain
with mobile electron\cite{galis}, the Ising-Heisenberg ladder model,
which is equivalent to the mixed spin-(3/2, 1/2) Ising-Heisenberg
diamond chain\cite{souza} and the spin-1/2 Ising-Heisenberg triangular
tube chain\cite{alecio}.

Inspired by the above results, we will investigate the thermal entanglement
and entropy behavior around the quasi-phase region for a spin-1/2
Ising-XYZ diamond chain. Although this model has already been studied
in previous works, where some of the authors investigated properties
such as magnetization, quantum entanglement, and magnetocaloric effect\cite{rojas,rojas-1}.
Here we focus on the pseudo-transition neighborhood, which exhibit
unexpected effects.

This paper is organized as follows: In Section II, we briefly review
the model. Then the leading results of thermal entanglement and entropy
behaviors are discussed in detail in the neighborhood of the quasi-phase
region in Section III. Finally, in Section IV, we summarize our conclusions.

\section{Hamiltonian of the model}

\begin{figure}
\includegraphics[scale=0.3]{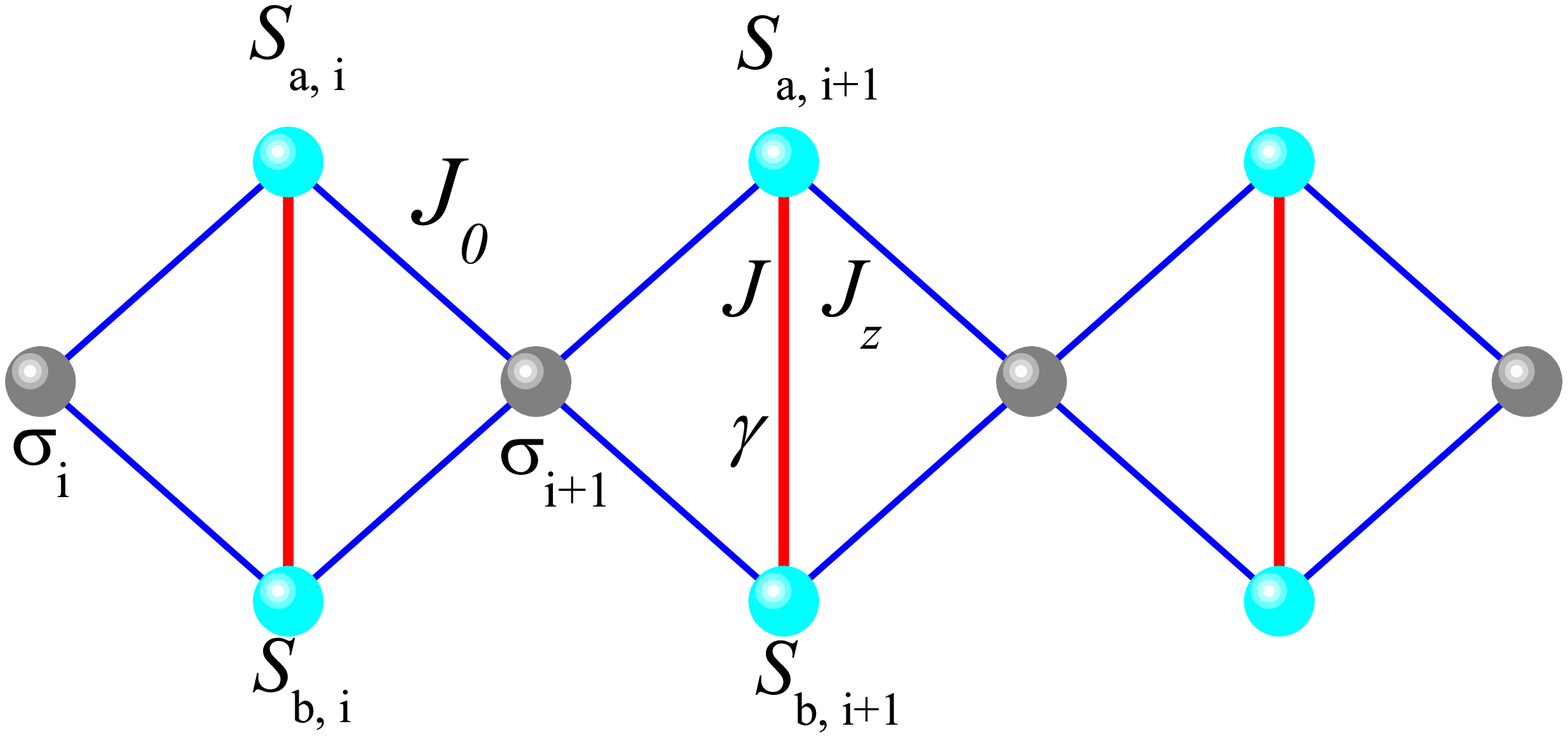} \caption{\label{fig:chain} Schematic representation of a spin-1/2 diamond
chain with Ising spins $\sigma$ localized in the nodal sites, and
Heisenberg spins $S$ localized in the interstitial sites. The coupling
parameters are $J_{0}$, $J_{z}$ and $J$, whereas $\gamma$ describes
the $xy$-anisotropy.}
\end{figure}

Let us consider the Hamiltonian of the Ising-XYZ diamond chain as
a sum of the Hamiltonian per unit cell $\mathcal{H}=\sum_{i=1}^{N}H_{i}$.
This Hamiltonian already was discussed in reference \cite{rojas},
which is described schematically in Fig.\ref{fig:chain}. Here we
rewrite this Hamiltonian as: 
\begin{alignat}{1}
H_{i}= & -J(1+\gamma)S_{a,i}^{x}S_{b,i}^{x}-J(1-\gamma)S_{a,i}^{y}S_{b,i}^{y}\nonumber \\
 & -J_{z}S_{a,i}^{z}S_{b,i}^{z}-J_{0}(S_{a,i}^{z}+S_{b,i}^{z})(\sigma_{i}+\sigma_{i+1})\nonumber \\
 & -h(S_{a,i}^{z}+S_{b,i}^{z})-\frac{h}{2}(\sigma_{i}+\sigma_{i+1}),\label{eq:Ham}
\end{alignat}
where $S_{a(b)}^{\alpha}(\alpha=x,y,z)$ denotes the Heisenberg spins
and $\sigma_{i}$ corresponds to the Ising spins. Whereas $\gamma$
means the $xy$-anisotropy, $J_{0}$ represents the Ising-type interaction
between nodal and interstitial sites, $J$ and $J_{z}$ are the Heisenberg-type
interaction between interstitial sites. We assume the system is subject
to an external magnetic field $h$ along the $z-$axis.

The eigenvalues of the Hamiltonian (\ref{eq:Ham}) for the $i$-th
unit cell are given by: 
\begin{eqnarray}
e_{1} & = & -h\frac{\mu}{2}-\frac{J_{z}}{4}+\Delta(\mu),\\
e_{2} & = & -h\frac{\mu}{2}-\frac{J}{2}+\frac{J_{z}}{4},\\
e_{3} & = & -h\frac{\mu}{2}+\frac{J}{2}+\frac{J_{z}}{4},\\
e_{4} & = & -h\frac{\mu}{2}-\frac{J_{z}}{4}-\Delta(\mu),
\end{eqnarray}
where $\mu=\sigma_{i}+\sigma_{i+1}$ and $\Delta(\mu)=\sqrt{(h+J_{0}\mu)^{2}+\frac{1}{4}J^{2}\gamma^{2}}$.
The corresponding eigenstates of the Hamiltonian in the standard basis
$\{|\begin{smallmatrix}+\\
+
\end{smallmatrix}\rangle,|\begin{smallmatrix}+\\
-
\end{smallmatrix}\rangle,|\begin{smallmatrix}-\\
+
\end{smallmatrix}\rangle,|\begin{smallmatrix}-\\
-
\end{smallmatrix}\rangle\}$ are:

\begin{eqnarray}
|\varphi_{1}\rangle & = & \frac{1}{\sqrt{1+\alpha_{+}^{2}}}\left(\alpha_{+}|\begin{smallmatrix}+\\
+
\end{smallmatrix}\rangle+|\begin{smallmatrix}-\\
-
\end{smallmatrix}\rangle\right),\\
|\varphi_{2}\rangle & = & \frac{1}{\sqrt{2}}\left(|\begin{smallmatrix}-\\
+
\end{smallmatrix}\rangle+|\begin{smallmatrix}+\\
-
\end{smallmatrix}\rangle\right),\\
|\varphi_{3}\rangle & = & \frac{1}{\sqrt{2}}\left(|\begin{smallmatrix}-\\
+
\end{smallmatrix}\rangle-|\begin{smallmatrix}+\\
-
\end{smallmatrix}\rangle\right),\\
|\varphi_{4}\rangle & = & \frac{1}{\sqrt{1+\alpha_{-}^{2}}}\left(\alpha_{-}|\begin{smallmatrix}+\\
+
\end{smallmatrix}\rangle+|\begin{smallmatrix}-\\
-
\end{smallmatrix}\rangle\right),\label{eq:estados}
\end{eqnarray}
where $\alpha_{\pm}=\frac{-J\gamma}{2h+2J_{0}\mu\pm2\Delta(\mu)}$.

\subsection{Ground states and quasi-phases}

\begin{figure}
\includegraphics[scale=0.25]{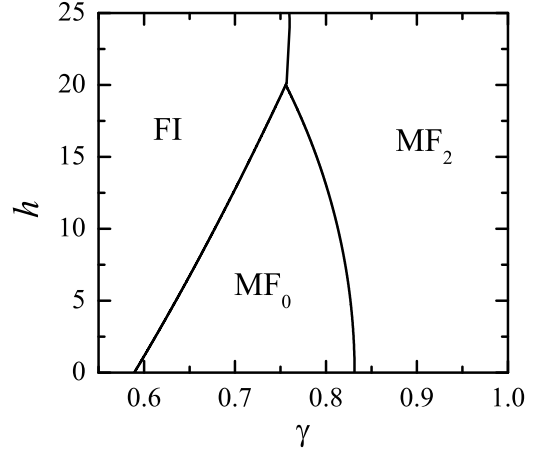} \caption{\label{diagrama} Phase diagram in the plane $\gamma$-$h$ at zero
temperature, for fixed parameters $J=100$, $J_{0}=-24$ and $J_{z}=24$. }
\end{figure}

First, let us analyze the phase diagram at zero temperature of the
Ising-XYZ diamond chain. We can observe that the phase diagram, illustrated
in Fig.\ref{diagrama} displays three phases: the eigenstates describe
one ferrimagnetic phase ($FI$) and two modulated ferromagnetic phases
($MF_{0}$) and ($MF_{2}$), these phases are: 
\begin{eqnarray}
\left|FI\right\rangle = & \overset{N}{\underset{i=1}{\prod}}|\varphi_{2}\rangle_{i}\otimes|+\rangle_{i},\label{eq:state4}\\
\left|MF_{0}\right\rangle = & \overset{N}{\underset{i=1}{\prod}}|\varphi_{4}\rangle_{i}\otimes|-\rangle_{i},\label{eq:state2}\\
\left|MF_{2}\right\rangle = & \overset{N}{\underset{i=1}{\prod}}|\varphi_{4}\rangle_{i}\otimes|+\rangle_{i}.\label{eq:state3}
\end{eqnarray}
Explicitly, the above states can be described as follows: the $FI$
phase corresponds to the antiferromagnetic Heisenberg spins with ferromagnetically
oriented Ising spins upwards, the $MF_{0}$ phase corresponds to modulated
ferromagnetic Heisenberg spins with Ising spins oriented ferromagnetically
to downwards, $MF_{2}$ phase corresponds to modulated ferromagnetic
Heisenberg spins with Ising spins oriented ferromagnetically upwards.

Whereas the corresponding ground state energies are: 
\begin{eqnarray}
\varepsilon_{2,1} & =E_{FI}= & \dfrac{J_{z}}{4}-\dfrac{h}{2}-\dfrac{J}{2},\label{eq:Eps_21}\\
\varepsilon_{0,0} & =E_{MF_{0}}= & -\dfrac{J_{z}}{4}+\dfrac{h}{2}-\sqrt{(h-J_{0})^{2}+\frac{1}{4}J^{2}\gamma^{2}},\label{eq:Eps_00}\\
\varepsilon_{2,0} & =E_{MF_{2}}= & -\dfrac{J_{z}}{4}-\dfrac{h}{2}-\sqrt{(h+J_{0})^{2}+\frac{1}{4}J^{2}\gamma^{2}}.\label{eq:Eps_20}
\end{eqnarray}
We observe in Fig.\ref{diagrama} that the interface between the $MF_{0}$
and $MF_{2}$ phases, $FI$ and $MF_{0}$ phases are non-degenerate,
this is one of the key reasons for the rise of pseudo-transition.
Whereas, the phase transition boundary between $FI$ and $MF_{2}$
phases is frustrated ($2^{N}$ degenerate) with a residual entropy
per unit cell $\mathcal{S}=\ln(2)$.

\subsection{Thermodynamics }

Here we give a brief review concerning the thermodynamics of the model,
the partition function is written as: 
\begin{eqnarray}
\mathcal{Z}=\sum_{\{\sigma\}}{\rm {e}^{-\beta\mathcal{H}},}
\end{eqnarray}
where $\beta=1/k_{B}T$, $k_{B}$ is the Boltzmann constant and $T$
is the absolute temperature. Through the manuscript we will consider
in units of $k_{B}$, or simply set as $k_{B}=1$.

Using the transfer matrix approach\cite{baxter} the solution for
the Ising-XYZ diamond chain was obtained in reference \cite{rojas}.
The transfer matrix eigenvalues are given by: 
\begin{eqnarray}
\lambda_{\pm}=\frac{w_{1}+w_{-1}\pm\sqrt{(w_{1}-w_{-1})^{2}+4w_{0}^{2}}}{2},
\end{eqnarray}
with 
\begin{eqnarray*}
w_{\mu}=2\mathrm{e}^{\frac{\beta\mu h}{2}}\left[\mathrm{e}^{-\frac{\beta J_{z}}{4}}\cosh\left(\frac{\beta J}{2}\right)+\mathrm{e}^{\frac{\beta J_{z}}{4}}\cosh\left(\beta\Delta_{\mu}\right)\right],
\end{eqnarray*}
are the elements of the transfer matrix.

The free energy per unit cell, in the thermodynamic limit $N\rightarrow\infty$,
becomes: 
\begin{eqnarray}
f=-\lim_{N\rightarrow\infty}\frac{1}{\beta N}\ln\mathcal{Z}=-\frac{1}{\beta}\ln\lambda_{+}.\label{free}
\end{eqnarray}
In the next section, we show our analyzes around the quasi-phase regions.

\subsection{Quantum entanglement}

The concurrence $\mathcal{C}$ can quantify the thermal entanglement
of a pair Heisenberg spins for same unit cell according to the reference
\cite{wootters}, which is defined as: 
\begin{eqnarray}
\mathcal{C}={\rm {max}\left\{ 0,\sqrt{\Lambda_{1}}-\sqrt{\Lambda_{2}}-\sqrt{\Lambda_{3}}-\sqrt{\Lambda_{4}}\right\} \label{31},}
\end{eqnarray}
where $\Lambda_{i}\:(i=1,2,3,4)$ are the eigenvalues in decreasing
order associated with the matrix: 
\begin{eqnarray}
R=\rho\left(\sigma^{y}\otimes\sigma^{y}\right)\rho^{\ast}\left(\sigma^{y}\otimes\sigma^{y}\right),\label{32}
\end{eqnarray}
with $\sigma^{y}$ being the Pauli matrix and $\rho^{\ast}$ denotes
complex conjugation of the density operator $\rho$. The concurrence
can be expressed in terms of the elements of $\rho$, 
\begin{eqnarray}
\mathcal{C}=2{\rm {max}\left\{ 0,\mid\rho_{14}\mid-\sqrt{\rho_{22}\rho_{33}},\mid\rho_{23}\mid-\sqrt{\rho_{11}\rho_{44}}\right\} \label{33},}
\end{eqnarray}
where the elements of the reduced density matrix $\rho_{ij}$ was
expressed in terms of the correlation functions, as shown explicitly
in reference \cite{amico}.

\begin{eqnarray}
\rho_{11} & =\frac{1}{4}+\langle S_{a}^{z}S_{b}^{z}\rangle+\langle S_{a}^{z}\rangle,\\
\rho_{22} & =\rho_{33}=\frac{1}{4}-\langle S_{a}^{z}S_{b}^{z}\rangle,\\
\rho_{44} & =\frac{1}{4}+\langle S_{a}^{z}S_{b}^{z}\rangle-\langle S_{a}^{z}\rangle,\\
\rho_{14} & =\langle S_{a}^{x}S_{b}^{x}\rangle-\langle S_{a}^{y}S_{b}^{y}\rangle,\\
\rho_{23} & =\langle S_{a}^{x}S_{b}^{x}\rangle+\langle S_{a}^{y}S_{b}^{y}\rangle.
\end{eqnarray}
Using this result, we can study the thermal entanglement around the
quasi-phases regions.

\section{results and discussions}

In what follows we present and discuss the results concerning the
quantum entanglement for a pair Heisenberg spins and entropy per unit
cell, in the neighborhood of pseudo-transition.

\begin{figure}
\includegraphics[scale=0.1]{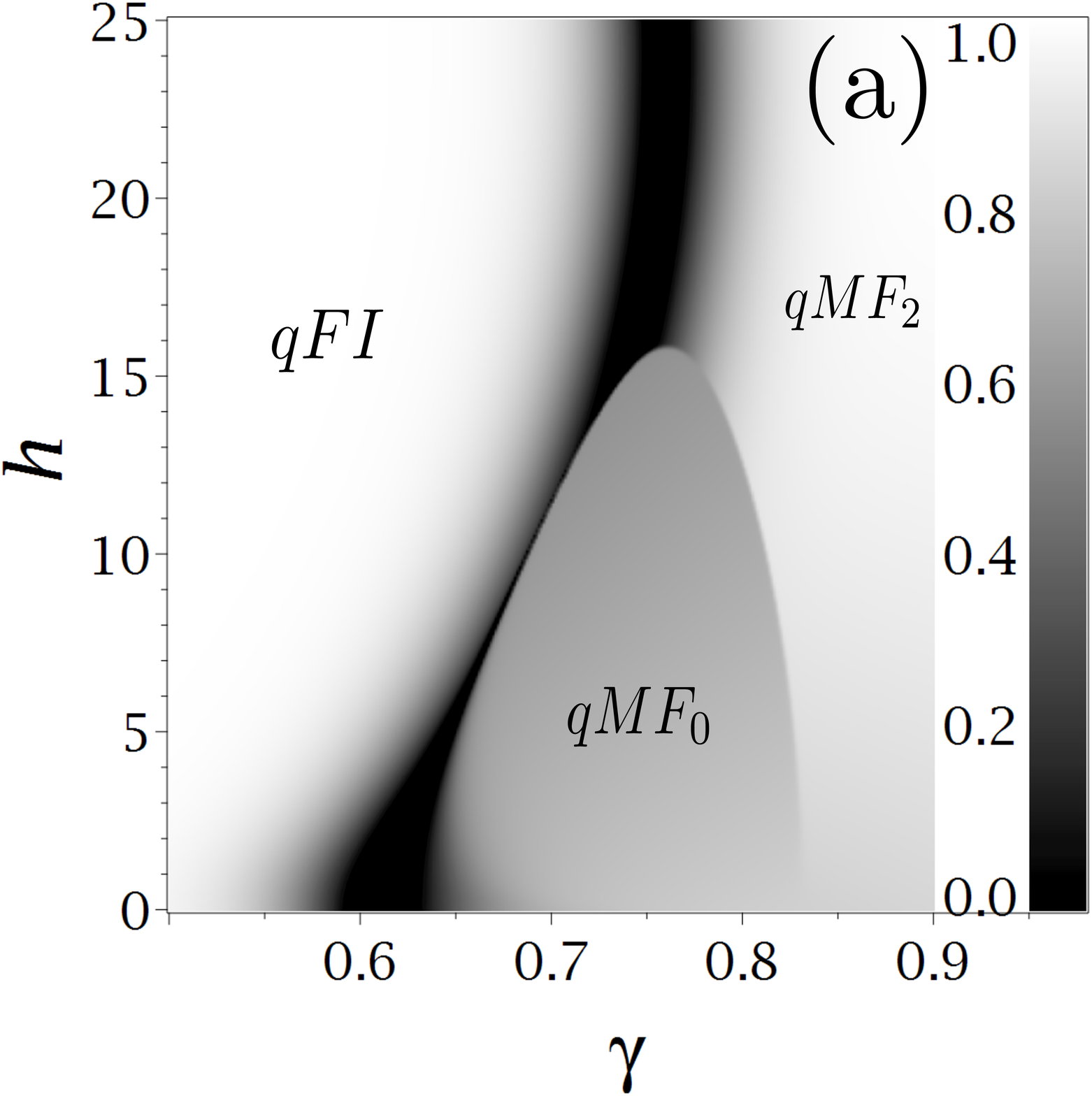}\includegraphics[scale=0.1]{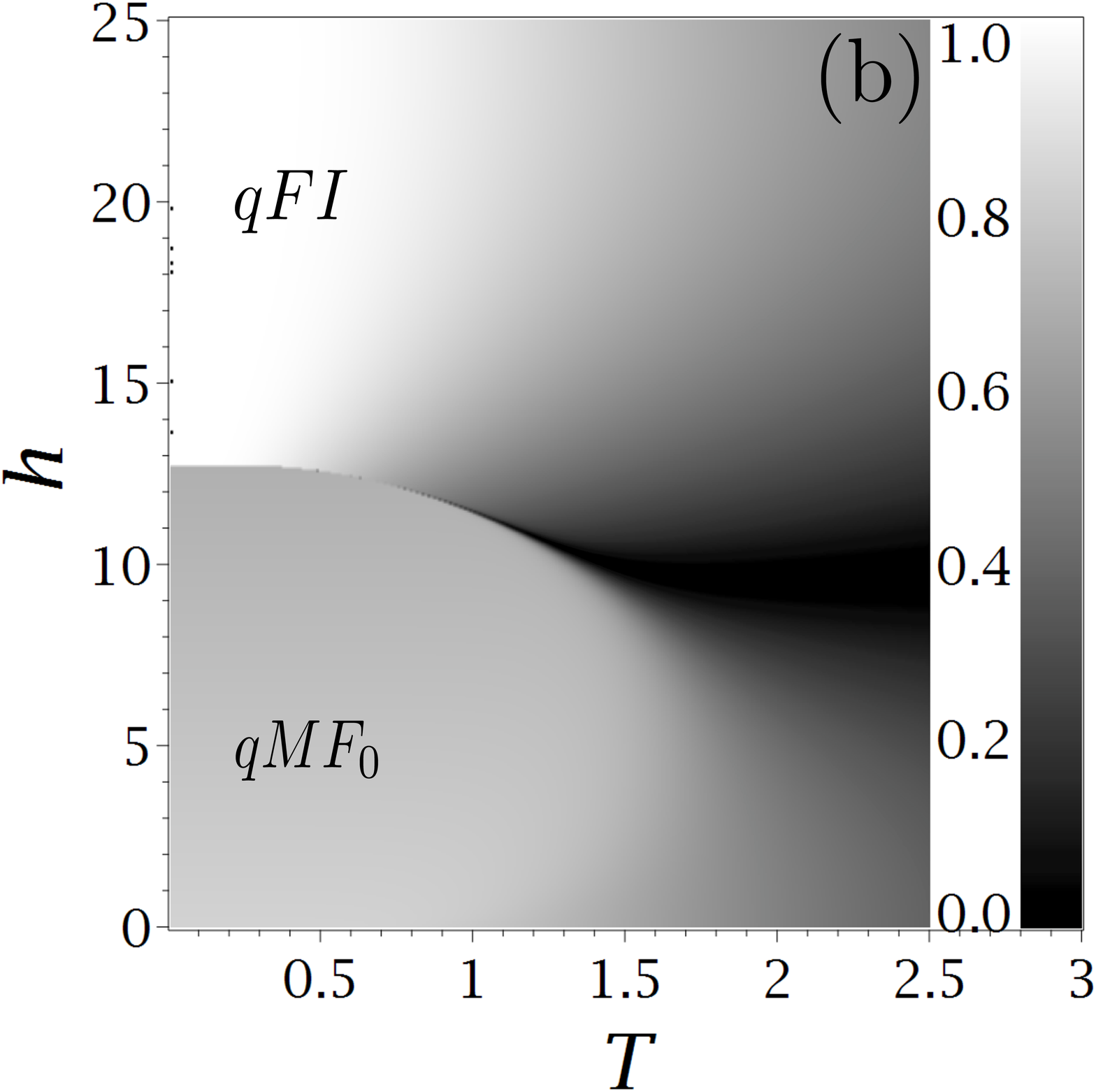}
\caption{\label{fig:1} Density plot of the concurrence. In (a) we illustrate
the concurrence as a function of magnetic field $h$ and $xy$-anisotropy
$\gamma$ for a fixed value of the temperature $T=1$. In (b) shows
the concurrence as a function of magnetic field $h$ and temperature
$T$ for a fixed $\gamma=0.7$. Assuming established parameters $J=100$,
$J_{0}=-24$ and $J_{z}=24$. }
\end{figure}

In Fig.\ref{fig:1}a we illustrate the concurrence as a function of
magnetic field $h$ and the $xy$-anisotropy $\gamma$, for fixed
temperature $T=1$ considering the spin parameters in the legend of
Fig.\ref{fig:1}. We extend the zero temperature phases to finite
temperature, where the phase becomes quasi-phases (compare Fig.\ref{diagrama}
and Fig.\ref{fig:1}a): $FI\rightarrow qFI$, $MF_{0}\rightarrow qMF_{0}$
and $MF_{2}\rightarrow qMF_{2}$. These quasi-phases mean that the
spins configurations mostly remains in the ground states configurations
given by (\ref{eq:state4}), (\ref{eq:state2}) and (\ref{eq:state3}).
We observe between the $qFI$ and $qMF_{0}$ occurs a pseudo-transition,
and this interface for a pair spins has null entanglement. Although
the interface between $FI$ and $MF_{0}$ states for a pair spins
is entangled at $T=0$. In the low-temperature limit, this interface
is strongly influenced by the interface $qFI$ and $qMF_{2}$ which
is a frustrated state, with null entanglement at $T=0$. However,
the interface between the quasi-phase $qMF_{0}$ and $qMF_{2}$ for
a pair spins are entangled, because the interface between $qMF_{0}$
and $qMF_{2}$ for a pair spins are entangled in $T=0$.

In Fig.\ref{fig:1}b is shown the concurrence as a function of the
magnetic field $h$ and temperature $T$. The pair spins in the region
$qFI$ is maximally entangled at low temperature, and the entanglement
decreases when increases the temperature, presenting a standard behavior.
Nevertheless, in the region $qMF_{0}$ we can notice that entanglement
is almost constant up to a temperature $T\approx1.5$ and after this
temperature the behavior of entanglement becomes standard as in ordinary
chain model.

\begin{figure*}
\includegraphics[scale=0.6]{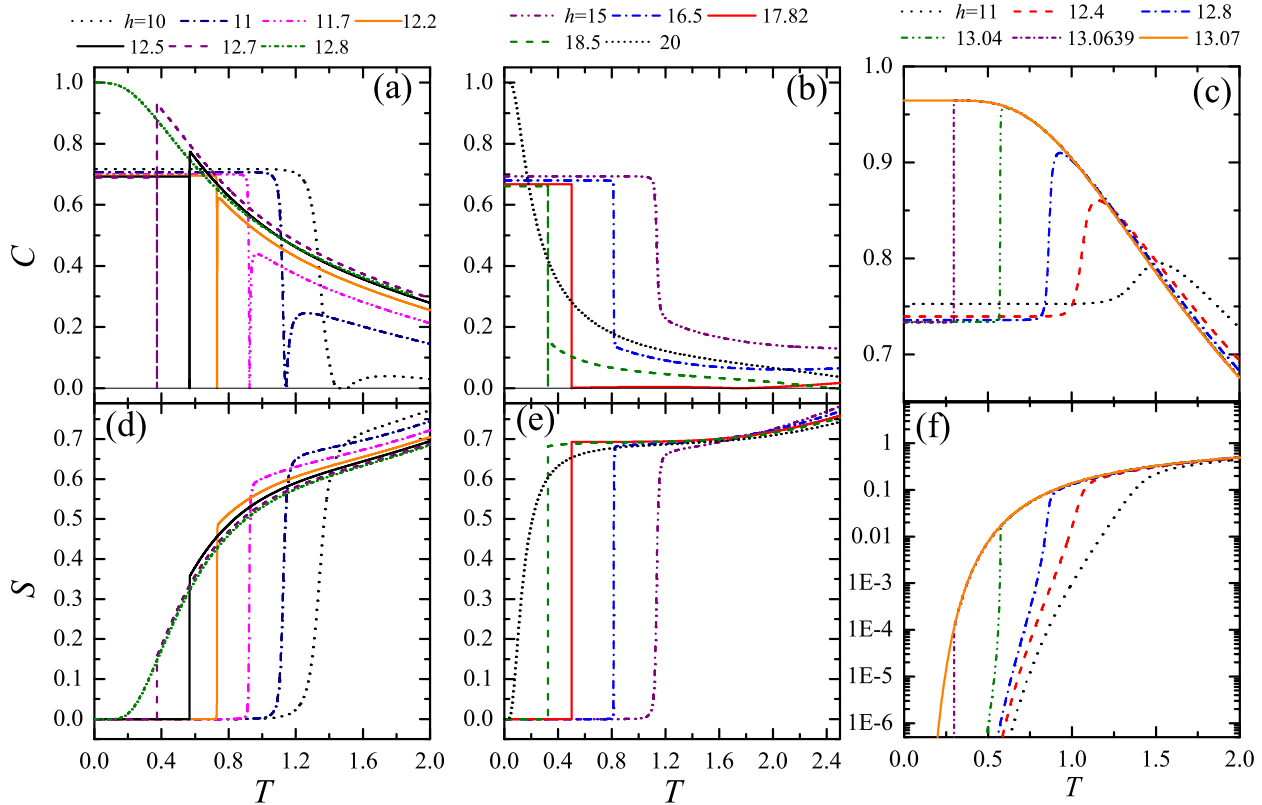}

\caption{\label{fig:2} Concurrence $\mathcal{C}$ as a function of the temperature
$T$ (a-c) and entropy $\mathcal{S}$ versus temperature $T$ (d-f),
fixed $J=100$, $J_{0}=-24$, $J_{z}=24$. In first column for $\gamma=0.7$,
in the second column for $\gamma=0.75$ and in the third column for
$\gamma=0.8$.}
\end{figure*}

The concurrence $\mathcal{C}$ is shown versus temperature $T$ in
Fig.\ref{fig:2}(a-c) for different values of the external magnetic
field $h$ and $\gamma$: first column is for $\gamma=0.7$, second
column is for $\gamma=0.75$ and third column is for $\gamma=0.8$.
The quantum entanglement presents an unusual behavior, which remains
almost constant up to pseudo-critical temperature. The other interesting
behavior is in the pseudo-transition (Fig.\ref{fig:2}(a-b)), where
the entanglement is null between the pseudo-transition $qFI$ and
$qMF_{0}$, due to the strong contribution of $qMF_{2}$.

In Fig.\ref{fig:2}a we can observe that for magnetic field $h\geq12.8$
the concurrence has a standard behavior. For magnetic field $h\leq12.75$
the concurrence, remains almost constant $\mathcal{C}\approx0.6885$
up to pseudo-critical temperature $T_{p}<0.3726$, but in pseudo-critical
temperature, the concurrence drops sharply to zero, due the spin systems
is disentangled in the frustrated interface between $qFI$ and $qMF_{2}$
states. For magnetic field $h=12.5$, we can notice the same behavior
as for $h=12.7$. For $h=10$ and $T_{p}\leq1.4369$ the concurrence
persist almost constant and decreasing to zero less abruptly, since
the influence of interface $qFI$ and $qMF_{2}$ is suppressed by
thermal excitation. The concurrence remains zero in the interval of
$1.4369<T_{p}<1.5103$, for higher temperature the concurrence increases
up to $\mathcal{C}\approx0.0374$ and then decreases.

In Fig.\ref{fig:2}b we observe a similar behavior to that reported
in (a). In Fig.\ref{fig:2}c shows the concurrence for magnetic field
$h=13.0639$. In this case, the concurrence is almost constant up
to the pseudo-critical temperature $T_{p}\approx0.2994$, for higher
temperature, the concurrence jumps from $\mathcal{C}\approx0.733515$
to $\mathcal{C}\approx0.9644$. Because the concurrence in $qMF_{0}$
at zero temperature is $\mathcal{C}=0.73364$ and in the concurrence
in $qMF_{2}$ region is $\mathcal{C}=0.9647$ at $T=0$. The same
behavior occurs for a lower magnetic field. For magnetic fields smaller
than $h<13.07$ the concurrence leads to standard behavior because
there is no pseudo-transition.

We also analyzed the behavior of entropy per unit cell $\mathcal{S}=-(\partial f/\partial T)$
in pseudo-transition as a function of the temperature $T$. In low
temperature limit ($T<T_{p}$) the entropy is given by 
\begin{equation}
\mathcal{S}\approx(1-\frac{2\Delta}{T}){\rm e}^{-2\Delta/T},\label{eq:Entrop-lw}
\end{equation}
where $\Delta={\rm min}(\varepsilon_{1,0}-\varepsilon_{0,0},\frac{\varepsilon_{0,1}-\varepsilon_{0,0}}{2})$
and $\varepsilon_{0,0}$ is given in \eqref{eq:Eps_00}, whereas the
other energies were defined in reference \cite{qphases} $\varepsilon_{0,1}=J_{z}/4-(J-h)/2$,
$\varepsilon_{1,0}=J_{z}/4-J/2$. When $\Delta/T\gg1$ the entropy
leads to $\mathcal{S}\rightarrow0$. 

In Fig.~\ref{fig:2}d is illustrated that the entropy as a function
of temperature, where we observe the entropy remains almost null given
by \eqref{eq:Entrop-lw}, presenting an abrupt jump at pseudo-critical
temperature $T_{p}$, after the jump the entropy increases smoothly
with increasing the temperature. In Fig.~\ref{fig:2}e is illustrated
the entropy as a function of the temperature, where we observe the
entropy is null bellow $T_{p}$ and above $T_{p}$ (but close to $T_{p}$)
the curve is almost a constant leading to $\mathcal{S}=\ln(2)$, whereas
for higher temperature the system behaves as for ordinary one. We
can note when the entropy reaches the pseudo-critical temperature
it jumps up to $\mathcal{S}\approx\ln(2)$ because the interface between
$qFI$ and $qMF_{2}$ is a frustrated region due to the Ising spins
configuration and presents a residual entropy of $\mathcal{S}=\ln(2)$
in $T=0$. In Fig.~\ref{fig:2}(f) we illustrate the entropy $\mathcal{S}$
versus $T$, where we choose a logarithmic scale because in this scale
we magnify the small entropy behavior, observing still a small jump
in entropy.

\section{Conclusion}

Despite there is no real phase transition in one-dimensional models.
Lately, have been explored some peculiar one-dimensional models that
manifest a surprising pseudo-transitions and quasi-phases. This pseudo-transition
closely follows a first- and second-order phase transition simultaneously.
So we consider a spin-1/2 Ising-XYZ diamond chain with external magnetic
field applied along the $z$-axis, where the Ising spins are located
at the nodal sites, and the Heisenberg spins are located at the interstitial
sites, with Ising-type interaction between the nodal and interstitial
sites, and Heisenberg-type interaction between interstitial sites.
This model has already been extensively studied in the literature,
the particular case we consider here exhibits a fascinating property
as the pseudo-transition. We explore how this property manifests in
quantum entanglement and entropy, near the pseudo-transition, as well
as in quasi-phase regions ($qFI$, $qMF_{0}$ and $qMF_{2}$). We
study the quantum entanglement by measuring the concurrence. We observe
that the temperature-dependent concurrence in the region $qMF_{0}$
remains almost constant up to a pseudo-critical temperature. By analyzing
the quantum entanglement at the interface of $qFI$ and $qMF_{0}$,
we find that quantum entanglement is null, due to the influence of
the interface between $qFI$ and $qMF_{2}$ which is frustrated state
with null entanglement. We analyze the entropy as a function of temperature
and observe an unusual behavior of the entropy that occurs when it
reaches the pseudo-critical temperature. We can also observe that
the entropy remains almost null with the temperature, up to the pseudo-critical
temperature, after which the entropy jumps and then behaves like a
standard spin chain. We find that at the interface between $qFI$
and $qMF_{2}$, the entropy upon reaching the pseudo-critical temperature
jumps from almost zero to $\mathcal{S}\approx\ln(2)$, and remains
practically constant and slowly starts to increase, this is because
the interface is frustrated at $T=0$ with residual entropy per unit
cell $\mathcal{S}=\ln(2)$.

\section*{Acknowledgments}

O. Rojas, M. Rojas and S. M. de Souza thank CNPq, CAPES and FAPEMIG
for partial financial support. J. Torrico and I. M. Carvalho thanks
CAPES and CNPq for full financial support.

\end{document}